\documentclass[aps,prd,twocolumn, showpacs, floatfix, letterpaper,preprintnumbers,superscriptaddress,nofootinbib]{revtex4-2}

\pdfoutput=1 
\usepackage[utf8]{inputenc}
\usepackage{amsmath}
\usepackage{amsfonts}
\usepackage{amssymb}
\usepackage{mathtools}
\usepackage{bm}
\usepackage{xcolor}
\usepackage{soul}
\usepackage{natbib}

\begin{document}

\newcommand{\vect}[1]{\boldsymbol{\mathbf{#1}}}

\setlength{\parskip}{0.08mm}

\newcommand{\order}[2]{\overset{\mathclap{\scriptscriptstyle #2}}{#1}\vphantom{#1}}
\newcommand{\lc}[1]{\overset{\circ}{#1}\vphantom{#1}}
\newcommand{\dd}{\mathrm{d}}

\newcommand{\etal}{\textit{et al.}}
\newcommand{\be}{\begin{equation}}
\newcommand{\ee}{\end{equation}}
\newcommand{\bea}{\begin{eqnarray}}
\newcommand{\eea}{\end{eqnarray}}
\renewcommand{\baselinestretch}{1.02}
\newcommand{\imsize}{0.83\columnwidth}
\newcommand{\halfsize}{0.38\columnwidth}
\newcommand{\bra}[1]{\left\langle #1\right|}
\newcommand{\ket}[1]{\left| #1\right\rangle}
\newcommand{\vev}[1]{\left\langle #1\right\rangle}
\newcommand{\avg}[1]{\langle #1\rangle}
\newcommand{\half}{{\textstyle{1\over 2}}}
\newcommand{\lbar}[1]{\overline{#1}}
\newcommand{\vsone}{\vspace{1cm}}
\newcommand{\nn}{\nonumber}
\newcommand{\nd}[1]{/\hspace{-0.6em} #1}
\newcommand{\nk}{\noindent}
\newcommand{\norm}[1]{\lVert#1\rVert}
\newcommand{\abs}[1]{\lvert#1\rvert}
\def\DerN#1{\frac{d #1}{d N}}
\def\rf#1{(\ref{#1})}
\newcommand{\reef}[1]{(\ref{#1})}
\def\de{\partial}
\def\C{{\cal{C}}}
\def\l{\langle}
\def\r{\rangle}


\title{Parametrized post-Newtonian formalism in higher-order Teleparallel Gravity}
\author{Manuel Gonzalez-Espinoza}
\email{manuel.gonzalez@pucv.cl}
\affiliation{Instituto de F\'{\i}sica, Pontificia Universidad Cat\'olica de
Valpara\'{\i}so, Casilla 4950, Valpara\'{\i}so, Chile}

\author{Giovanni Otalora}
\email{giovanni.otalora@pucv.cl}
\affiliation{Instituto de F\'{\i}sica, Pontificia Universidad Cat\'olica de
Valpara\'{\i}so, Casilla 4950, Valpara\'{\i}so, Chile}
 
\author{Lucila Kraiselburd}
\email{lkrai@fcaglp.unlp.edu.ar}
\affiliation{Facultad de Ciencias Astron\'omicas y Geof\'isicas, Universidad Nacional de La Plata, Paseo del Bosque S/N, C.P. 1900, La Plata, Argentina}
\affiliation{CONICET, Godoy Cruz 2290, C.P. 1425, Ciudad Aut\'{o}noma de Buenos Aires, Argentina.}
        
\author{Susana Landau}
\email{slandau@df.uba.ar}
\affiliation{Departamento de F\'{\i}sica FCEN-UBA and IFIBA CONICET-UBA, 
        Facultad de Ciencias Exactas y Naturales, 
         Universidad de Buenos Aires , 
        Ciudad Universitaria - Pab. I,\\ 
        C.P. 1428, Ciudad Aut\'{o}noma de Buenos Aires,
        Argentina}
\affiliation{CONICET, Godoy Cruz 2290, C.P. 1425, Ciudad Aut\'{o}noma de Buenos Aires, Argentina.}

\begin{abstract} 
 We study the parametrized post-Newtonian (PPN) limit of higher-derivative-torsion Modified 
Teleparallel Gravity. We start from the covariant formulation of modified Teleparallel Gravity by restoring the spin connection of the theory. Then, we perform the post-Newtonian expansion of the tetrad field around the Minkowski background and find the perturbed field equations. We compute the PPN metric for the higher-order Teleparallel Gravity theories  which allows us to show that at the post-Newtonian limit this more general class of theories are fully conservative and indistinguishable from General Relativity . In this way, we extend the results that were already found for $F(T)$ gravity in previous works. 
Furthermore,  our calculations reveal the importance of considering  a second post-Newtonian (2PN) order approximation or a parametrized post-Newtonian cosmology (PPNC) framework where additional perturbative modes coming from general modifications of Teleparallel Gravity could lead to new observable imprints.
\end{abstract}

\pacs{04.50.Kd, 98.80.-k, 95.36.+x}

\maketitle

\section{Introduction}\label{Introduction}
The discovery of the current accelerated expansion of the universe has led the community of theoretical physicists to pose a challenging problem that has not been solved yet: the explanation of the physical mechanism responsible for this phenomenon. 
The solution provided by the Standard Cosmological Model ($\Lambda$CDM) is to include a cosmological constant in Einstein equations. However, the observational value of this constant can not be explained by the Standard Model of Particle Physics. Other proposals often named as "dark energy models" consist in adding extra degrees of freedom to the Standard Model of Particle Physics. On the other hand, other authors have analyzed the possibility that the extra degrees of freedom are added to the gravitational sector of the theory, which leads to assume alternative theories of gravity to General Relativity (GR). It is important to stress that all theoretical proposals mentioned before are able to explain  current cosmological observations such as those provided by type Ia supernovae, the Cosmic Microwave Background and Baryon Acoustic Oscillations just to mention the most relevant ones. 
Besides, it is well known that there is a discrepancy between the values of the Hubble constant obtained using Cosmic Microwave Background data and those calculated using type Ia supernovae explosions together with local distance calibrations \cite{2021arXiv210901161S}. Even though, there is no agreement  about the amount of the discrepancy (some authors claim there is a 4$\sigma$ discrepancy while others report only 2$\sigma$ or even no discrepancy \cite{2021arXiv210511461M,2021ApJ...919...16F,2021ApJ...908L...6R}), it is clear that it can not be explained in the context of the Standard Cosmological Model. As a consequence, alternative cosmological models, in particular, those based in alternative theories of gravity, 
that have been considered in the literature before \cite{2016RPPh...79d6902K,2016RPPh...79j6901C}, have now a growing interest.  In this work we will focus in those theories where the lagrangian of the gravitational sector is replaced by an arbitrary function of the torsion scalar, the so called $f(T) $ theories \cite{2016RPPh...79j6901C}. The torsion scalar is defined in the context of the so-called teleparallel equivalent of General Relativity or simply Teleparallel Gravity (TG) \cite{Einstein,TranslationEinstein,Early-papers1,Early-papers2,Early-papers3,Early-papers4,Early-papers5,Early-papers6,JGPereira2,AndradeGuillenPereira-00,Arcos:2005ec,Pereira:2019woq}. This is a gauge theory for the translation group where the dynamical variable is the tetrad field whose non-trivial part representing the gravitational field is the translational gauge potential and the field strength is the corresponding torsion tensor \cite{Aldrovandi-Pereira-book}. Then, unlike GR, in this alternative and equivalent description of gravity the Lorentz connection is purely inertial giving a vanishing curvature tensor \cite{Krssak:2015rqa}.   
While alternative theories of gravity are able to explain the current accelerated expansion of the universe without a cosmological constant, it has been pointed out  that some of them are ruled out with the the bounds imposed by local experiments such for example those performed within the solar system.

The $f (T)$ theories have already been studied in the context of the solar system using i) different effects such as perihelion precession, Shapiro time delay, gravitational redshift and light bending; and ii) appealing to the parametrized post-Newtonian (PPN)  formalism. This later approach provides a solution to the Einstein equations for different metric theories in the weak-field slow-motion limit, and generates ten parameters which can be compared with high precision solar system data to establish the viable regions of a theory. In this way, it is not necessary to calculate the theoretical predictions for each effect to compare them with the observations, but it is enough to estimate the theoretical PPN parameters and contrast them with the PPN values obtained from the observational data \cite{will1993,Will2014,Will2018}.  The first $f(T)$ published analyses using the solar system effects \cite{Xie2013,Iorio:2015rla,Ruggiero2015,Ruggiero2016,Farrugia:2016xcw}, were obtained  considering a bad choice of the tetrad which consequently  triggered an incorrect solution \footnote{The tetrads used in those articles do not yield  torsion scalars that vanish in the Minkowski spacetime limit.}. 
Later on,  this mistake was corrected using a covariant formulation of $f(T)$ Gravity \cite{Krssak:2015oua}, which allowed  the old results to be improved  \cite{DeBenedictis2016,Bahamonde2019,Otalora:2017qqc}. Moreover, a similar analysis was performed for $f(T,B)$ theories  ($f(T)$ theories  with a boundary term $B$)  \cite{Bahamonde2020}. On the other hand, several articles  have applied the PPN formalism to Modified Teleparallel Gravity (MTG) theories such a for example Ref \cite{Ualikhanova2019}, where the post-Newtonian limit of a general class of Teleparallel Gravity theories (that includes $f (T)$ theories) is derived by imposing the Weitzenb\"{o}ck gauge. Furthermore, assuming a post-Newtonian approximation of the tetrad around a Minkowski background solution; the authors concluded that   the $f(T)$ predictions in this limit are not distinguished from those of GR. This approach has also been used   in other works such as \cite{Flathman2020}, where Teleparallel Gravity theories whose action is a free function $L(T,X,Y,\phi)$ of the torsional scalar $T$ and scalar quantities $X$ and $Y$ formed from a massless scalar field $\phi$ are studied. Another examples are Ref.  \cite{Emtsova2020}, which is focused on massive and massless scalar fields; and Ref. \cite{Flathmann2021}, in which an analysis of symmetric Teleparallel Gravity theories (whose action is defined by a free function of the five parity-even scalars that
are quadratic in the non-metricity tensor) is carried out. In this paper we focus in the analysis of  teleparallel theories  with higher-derivative torsional terms in the action $F(T,(\nabla T)^2, \Box T)$ within the framework of the PPN formalism, which has not been performed so far.
 
The manuscript is organized as follows: in Section \ref{fT_model} we provide a brief description of the main aspects of  $f(T)$ theories, specifically  those with  higher derivative torsional terms. Next, we  review in Sec. \ref{sec:ppn} the parameterized post-Newtonian (PPN) formalism focusing  on the necessary modifications to describe  Teleparallel Gravity.  Using this latter formalism  we compute in Sec. \ref{sec:solution} the components of the tetrad which allows us to obtain  in Sec. \ref{sec:metpar} the metric for a generalized $F(T,(\nabla T)^2,\Box{T})$ gravity. The PPN parameters are then  obtained from the comparison of the latter with the standard PPN metric. Finally, in Section \ref{conclusion_f} we present our conclusions and discuss several aspects to improve in the PPN formalism applied to these kind of theories.


\section{$F(T,(\nabla T)^2,\Box{T})$ gravity}\label{fT_model} 


The Teleparallel Gravity (TG) is an alternative formulation of gravity equivalent to GR where the dynamical field is given by the tetrad $e^{A}_{~\mu}$, that sets up an 
orthonormal base for the tangent space at each point of a manifold \cite{Einstein,TranslationEinstein,Early-papers1,Early-papers2,Early-papers3,Early-papers4,Early-papers5,Early-papers6,JGPereira2,AndradeGuillenPereira-00,Arcos:2005ec,Pereira:2019woq}. Also, it is connected to the metric through the following relationship,
\begin{equation}
g_{\mu\nu}=\eta_{A B} e^{A}_{~\mu} 
e^{B}_{~\nu},
\label{metricvierbeinrel}
\end{equation}
where greek indice span the coordinate space and latin indices span the 
tangent space. By using a general Lorentz frame one can write the tetrad field as \cite{Pereira.book}
\be
e^{A}_{~\mu}=\partial_{\mu}{x^{A}}+\omega^{A}_{~B \mu} x^{B}+B^{A}_{~\mu},
\label{tetrad}
\ee 
where the inertial effects are embedded into the spin connection $\omega^{A}_{~B \mu}$ and the gravitational field is represented by the translational gauge potential $B^{A}_{~\mu}$. Then, the spin connection of TG is given by
\be
\omega^{A}_{~B \mu}=\Lambda^{A}_{~D}(x)\partial_{\mu}\Lambda_{B}^{~D}(x), 
\ee
with $\Lambda^{A}_{~D}(x)$ a local (point-dependent) Lorentz transformation. This is a purely (flat) spin connection and then  it gives a vanishing curvature tensor \cite{Pereira.book,Krssak:2015rqa}. On the other hand, in the presence of gravitation ($B^{A}_{~\mu}\neq 0$) the tetrad field \eqref{tetrad} leads to the non-zero torsion tensor



\begin{equation}
T^\rho_{\verb| |\mu\nu} \equiv e_{A}^{~\rho}
\left( \partial_\mu e^{A}_{~\nu}-\partial_{\nu} e^{A}_{~\mu} +\omega^{A}_{~B \mu} e^{B}_{~\nu}-\omega^{A}_{~B \nu}e^{B}_{~\mu}\right).
\end{equation}
Consequently, the torsion scalar $T$ represents the Lagrangian of the theory, and is built from the contractions of the torsion tensor such that \cite{Pereira.book,Maluf:2013}, 
\begin{equation}
\label{Tdefscalar}
T\equiv\frac{1}{4}
T^{\rho \mu \nu}
T_{\rho \mu \nu}
+\frac{1}{2}T^{\rho \mu \nu }T_{\nu \mu\rho}
-T_{\rho \mu}{}^{\rho }T^{\nu\mu}{}_{\nu}.
\end{equation}


In addition, motivated by the $f(R)$ theories in which the scalar of curvature $R$ is replaced by a function of itself \cite{DeFelice:2010},  simple torsion-modified theories of gravity \cite{Bengochea:2008gz,Linder:2010py}  have been developed. Noticeably, the Lagrangian of these theories is written as an arbitrary function $f (T)$.  Moreover, other alternative gravity theories where higher derivative torsional terms  such as $(\nabla T)^2$ and $\Box T$ are introduced have also been considered \cite{Otalora:2016,Capozziello:2020CQG,Bajardi:2021}.
In this article we focus on these kinds of theories where the actions  takes the following form
\begin{equation}
S=\frac{1}{2 \kappa^2}\int{ e 
F(T,\left(\nabla{T}\right)^2,\Box {T})} d^{4}x+S_{m}(e^{A}_{\rho},\Psi_{m}), 
\label{action0}
\end{equation} where $\kappa^2= 8 \pi G$ and the light speed $c$ is set to one. Also it follows that $\left(\nabla{T}\right)^2=\eta^{A B} e_A^{~\mu} e_B^{~\nu} \nabla_{\mu}{T}\nabla_{\nu}{T}=g^{\mu\nu} 
\nabla_{\mu}{T}\nabla_{\nu}{T}$ and $  \Box 
{T}= \eta^{A B} e_A^{~\mu} 
e_B^{~\nu}  \nabla_{\mu}\nabla_{\nu}{T}=g^{\mu\nu}\nabla_{\mu}{\nabla_{\nu}}{T}$, being $e= \det \left(e^A_{~\mu} \right)=\sqrt{-g}$.  

Rewriting action (\ref{action0}) based on these new parameters  $X_{1}\equiv \left(\nabla{T}\right)^2$, 
$X_{2}\equiv \Box {T}$,  $F_{T}\equiv \partial F/\partial T$ and $F_{,X_{a}}\equiv 
\partial F/\partial X_{a}$ (with $a=1,2$); and varying it with respect to the tetrad; the field equations can be obtained,
\begin{eqnarray}
E_{A}^{~\rho} &\equiv& \frac{1}{e}\partial_{\mu}\left(e F_{,T} e_{A}^{~\tau} S_{\tau}^{~\rho \mu}\right)-F_{,T} 
e_{A}^{~\tau} S_{\nu}^{~\mu \rho} T^{\nu}_{~\mu \tau}\nonumber\\
&&+F_{,T} e_{B}^{~\tau} S_{\tau}^{~\mu \rho} \omega^{B}_{~A \mu}+\frac{1}{4} 
e_{A}^{~\rho}F
\nonumber\\
&& 
+\frac{1}{4}\sum_{a=1}^{2}\Bigg\{F_{,X_{a}} \frac{\partial X_{a}}{\partial 
e^{A}_{~\rho}} -\frac{1}{
e}\Bigg[\partial_{\mu} \Bigg(e F_{,X_{a}} 
\frac{\partial{X_{a}}}{\partial{\partial_{\mu}{e^{A}_{~\rho}}}}\Bigg)\nonumber\\
&&
 -\partial_{\mu}\partial_{\nu}\Bigg( e F_{,X_{a}} \frac{\partial 
{X_{a}}}{\partial{\partial_{\mu}\partial_{\nu} e^{A}_{~\rho}}}\Bigg)\Bigg]\Bigg\}
\nonumber\\
&& - \frac{1}{4 e}\partial_{\lambda}\partial_{\mu}\partial_{\nu} \Bigg(e F_{,X_{2}} 
\frac{\partial{X_{2}}}{\partial \partial_{\lambda}\partial_{\mu}\partial_{\nu}{e^{A}_{~\rho}}}
\Bigg) \nonumber\\
&&+\frac{\kappa^2}{2} e_{A}^{~\tau}\,
  {\mathcal{T}^{(m)}}_{\tau}^{~\rho}=0.
\label{FEquations}
\end{eqnarray}
Here we define  $S_{\rho}^{~\mu\nu}\equiv 
\frac{1}{2}\left(K^{\mu\nu}_{~~\rho}+\delta^{\mu}_{\rho}\,T^{\tau \nu}_{
~~\tau}-\delta^{\nu}_{\rho}\,T^{\tau\mu}_{~~\tau}\right)$ as the "superpotential", and 
$K^{\mu\nu}_{~~\rho}\equiv 
-\frac{1}{2}\left(T^{\mu\nu}_{~~\rho}-T^{\nu\mu}_{~~\rho}-T_{\rho}^{~\mu\nu}\right)$   
the contortion tensor. Besides, the matter energy 
momentum tensor is given by, 
\begin{equation}
e_{A}^{~\tau}\,
  {\mathcal{T}^{(m)}}_{\tau}^{~\rho}
    \equiv   \frac{1}{e} 
\frac{\delta{{\mathcal S}_m}}{\delta{e^{A}_{\rho}}}.
 \label{4}
\end{equation}

The covariant derivative of the mattter action is null as long as its only coupling with gravity (that is, with the tetrad) is minimal and also, the Lagrangian of matter is diffeomorphism invariant. On the other hand, in a general coordinate basis the field equations can be written as
\bea
 E_{\mu \nu}&\equiv&  F_{,T} G_{\mu \nu}+S_{\mu \nu}^{~~~\sigma}\partial_{\sigma}{F_{,T}}+\frac{1}{4}g_{\mu \nu} (F-F_{,T}T)\nonumber\\
&+&\frac{1}{4}\sum_{a=1}^{2}\Bigg\{e^{A}_{~\mu} g_{\rho \nu} F_{,X_{a}} \frac{\partial X_{a}}{\partial 
e^{A}_{~\rho}} \nonumber\\
&-& \frac{1}{
e}\Bigg[e^{A}_{~\mu} g_{\rho \nu} \partial_{\sigma} \Bigg(e F_{,X_{a}} 
\frac{\partial{X_{a}}}{\partial{\partial_{\sigma}{e^{A}_{~\rho}}}}\Bigg)\nonumber\\
&-&
 e^{A}_{~\mu} g_{\rho \nu}\partial_{\sigma}\partial_{\delta}\Bigg( e F_{,X_{a}} \frac{\partial 
{X_{a}}}{\partial{\partial_{\sigma}\partial_{\delta} e^{A}_{~\rho}}}\Bigg)\Bigg]\Bigg\}
\nonumber\\
& -& \frac{1}{4 e} e^{A}_{~\mu} g_{\rho \nu}\partial_{\lambda}\partial_{\sigma}\partial_{\delta} \Bigg(e F_{,X_{2}} 
\frac{\partial{X_{2}}}{\partial\partial_{\lambda}\partial_{\sigma}\partial_{\delta}{e^{A}_{~\rho}}}
\Bigg) \nonumber\\
&+&\frac{\kappa^2}{2} {\mathcal{T}^{(m)}}_{\mu \nu} =0,
\label{F_E_SP}
\eea 
being $G^{\mu}_{~\nu}=e_{A}^{~\mu} G^{A}_{~\nu}$ the Einstein tensor, with $G_{A}^{~\mu}\equiv e^{-1}\partial_{\nu}\left(e e_{A}^{~\sigma} S_{\sigma}^{~\mu\nu}\right)-e_{A}^{~\sigma} T^{\lambda}_{~\rho \sigma}S_{\lambda}^{~\rho \mu}+e_{B}^{~\lambda} S_{\lambda}^{~\rho \mu}\omega^{B}_{~A \rho}+\frac{1}{4}e_{A}^{~\mu} T$ and  $E^{\mu}_{~\nu}=e_{A}^{~\mu} E^{A}_{~\nu}$. Although the action of Teleparallel Gravity is local Lorentz invariant \cite{Pereira.book}, the gravitational part of the action \eqref{action0} is not anymore \cite{Sotiriou:2010mv,Li:2010cg}. Then, the modified field equations   \eqref{F_E_SP} are not symmetric. Indeed, the superpotential tensor $S_{\mu \nu}^{~~~\sigma}$ is not symmetric in its lower indices, as well as the terms coming from the new higher-order derivative terms added to the action, and then $E_{\mu \nu}$ is also not symmetric. Thus, the antisymmetric part of \eqref{F_E_SP} constitutes a set of six equations for six additional degrees of freedom (DOF) due to violation of local Lorentz symmetry \cite{Gonzalez-Espinoza:2020azh} (see also Refs. \cite{Gonzalez-Espinoza:2018gyl,Gonzalez-Espinoza:2019ajd,Gonzalez-Espinoza:2021mwr,Gonzalez-Espinoza:2021qnv,Leyva:2021fuo} and references therein).


Finally, some of us \cite{Otalora:2016} pointed out before that  the higher-order derivatives in Eq. (\ref{FEquations})  might be generating Ostrogradsky ghosts. However, since the theory is not formulated in the Einstein frame, these terms could also be indicating the existence of extra degrees of freedom. Considering that there is not yet a transformation between the Jordan and Einstein frames for torsional modified gravity theories, this type of analysis is beyond the scope of this work.

\section{Post-Newtonian approximation}\label{sec:ppn}
In this section, we quickly summarize the parametrized post-
Newtonian formalism  that we are going to use in this work. The PPN formulation is a method that allows solving the Einstein's field equations of metric theories in the weak field limit and assuming slow motions in such a way that it is possible to compare the theoretical predictions with observations or experiments, for example those from the solar system \cite{Will2014,Will2018}. The main hypotheses of the PPN formalism are: i) matter behaves like a perfect low-speed fluid; ii) all relevant physical quantities in the solution of the gravitational field equations can be expanded in orders of the velocity $v^i=\frac{u^i}{u^0}$ of the
source matter. Several authors have already applied this formalism to different scalar torsion theories before \cite{Ualikhanova2019,Flathman2020,Emtsova2020,Flathmann2021}. Therefore, we will use several of their developments to carry out the expansion of important quantities (such as the tetrad) in the orders of the velocity.

First, we recall that the energy momentum tensor of a perfect fluid with rest energy
density $\rho$, specific internal energy $\Pi$, pressure $p$, and
four-velocity $u^\mu$, can be expressed as:
\begin{equation}\label{eqn:tmunu}
\mathcal{T}^{\mu\nu} = (\rho + \rho\Pi + p)u^{\mu}u^{\nu} + pg^{\mu\nu}\,,
\end{equation}
where the normalization of the four-velocity $u^\mu$ with the metric is such that the following relation is met: $u^\mu u ^\nu g_{\mu \nu} = -1$.  In Cartesian coordinates the diagonal tetrad representing the Minkowski background is a proper tetrad and then we can choose $\omega^{A}_{~B \mu}=0$ \cite{Krssak:2015oua}.
Next, we consider first an expansion of the tetrad field in Eq. \eqref{tetrad} around the Minkowski background as follows:
\begin{equation}\label{eqn:tetradexp}
e^A{}_{\mu} = \delta^A{}_{\mu} + \tau^A{}_{\mu} = \delta^A{}_{\mu} + \order{\tau}{1}^A{}_{\mu} + \order{\tau}{2}^A{}_{\mu} + \order{\tau}{3}^A{}_{\mu} + \order{\tau}{4}^A{}_{\mu} + \mathcal{O}(5)\,,
\end{equation}
being $\delta ^A{}_{\mu}$ = diag(1,1,1,1); and each term $\order{\tau}{n}^A{}_{\mu}$ is of order $O(n)\sim \vec v^n$. For our calculation, we only consider velocity orders up to the fourth order. Then, in order to study the PPN limit of the theory we can choose the following ansatz for the perturbed tetrad field
\bea
&& e^{\hat{0}}_{~\mu}=\left(1+\order{\tau}{2}^{\hat{0}}_{~0}+\order{\tau}{4}^{\hat{0}}_{~0},~~\order{\tau}{3}^{\hat{0}}_{~i}\right),\:\: e^{a}_{~\mu}=\left(\order{\tau}{3}^{a}_{~0},~~\delta^{a}_{~i}+\order{\tau}{2}^{a}_{~i}\right),\nonumber\\
&& e_{\hat{0}}^{~\mu}=\left(1-\order{\tau}{2}^{\hat{0}}_{~0}-\order{\tau}{4}^{\hat{0}}_{~0}+(\order{\tau}{2}^{\hat{0}}_{~0})^2,~-\delta^{i}_{a} ~\order{\tau}{3}^{a}_{~0}\right), \nonumber\\
&& e_{a}^{~\mu}=\left(-\delta^{i}_{a}~\order{\tau}{3}^{\hat{0}}_{~i},~\delta^{i}_{a}-\delta^{j}_{a}~\delta^{i}_{b}~\order{\tau}{2}^{b}_{~j}+\delta^{j}_{a} ~\delta^{k}_{c}~\delta^{i}_{b} ~\order{\tau}{2}^{c}_{~j}~\order{\tau}{2}^{b}_{~k}\right).
\eea 
The hat notation denotes time and spatial algebraic indices. The above ansatz introduces seventeen DOF, but only sixteen are going to be independent as usual in the tetrad formalism for gravity. Furthermore, although in the context of GR and TEGR, six of them are Lorentz gauge degrees recovering the usual ten DOF, in the case of modified teleparallel gravity theories these six additional modes are no longer Lorentz gauge degrees because local Lorentz violation \cite{Sotiriou:2010mv,Li:2010cg}.



Thus, using the expansion of Eq. (\ref{eqn:tetradexp}) we can obtain the perturbed metric around a flat background $\overset{0}{g_{\mu \nu }} = \eta_{\mu \nu }=\eta_{A B}\delta^{A}_{~~\mu}\delta^{B}_{~~\nu}$ as follows
\begin{eqnarray}\label{eqn:metricexp}
\order{g}{2}_{00} &=& 2\order{\tau}{2}_{00}\,, \quad
\order{g}{2}_{ij} = 2\order{\tau}{2}_{(ij)}\,, \quad \nonumber\\
\order{g}{3}_{0i} &=& 2\order{\tau}{3}_{(i0)}\,, \quad
\order{g}{4}_{00} = -(\order{\tau}{2}_{00})^2 + 2\order{\tau}{4}_{00}\,.
\end{eqnarray}
 
where we have introduced $\tau_{\mu\nu} = \delta^A{}_{\mu}\eta_{AB}\tau^B{}_{\nu}$ and
$\order{\tau}{n}_{\mu\nu} = \delta^A{}_{\mu}\eta_{AB}\order{\tau}{n}^B{}_{\nu}$. In addition, from that same expression, the torsion scalar reduces to

\begin{eqnarray}\label{T_equ_order}
T &=& - \partial_{i}\order{\tau}{2}_{j k} \partial_{k}\order{\tau}{2}_{(i,j)}+\partial_{i}\order{\tau}{2}_{j k} \partial_{j}\order{\tau}{2}_{[i, k]} -\partial_{0}\order{\tau}{2}_{i j} \partial_{0}\order{\tau}{2}_{(i , j)} \nonumber\\
&& +\partial_{i}\order{\tau}{2}_{j k} \partial_{i}\order{\tau}{2}_{(j , k)}-\partial_{i}\order{\tau}{2}_{i j} \partial_{k}\order{\tau}{2}_{k j}
\nonumber\\
&& +\partial_{0}\order{\tau}{2}_{i i} \partial_{0}\order{\tau}{2}_{j j}+2\partial_{i}\order{\tau}{2}_{j j} \partial_{i}\order{\tau}{2}_{0 0}-\partial_{i}\order{\tau}{2}_{j j} \partial_{i}\order{\tau}{2}_{k k}
\nonumber\\
&& -2\partial_{i}\order{\tau}{2}_{i j} \partial_{j}\order{\tau}{2}_{0 0}+2\partial_{i}\order{\tau}{2}_{i j} \partial_{j}\order{\tau}{2}_{k k} + \mathcal{O}(5),
\end{eqnarray}
where, $\order{\tau}{2}_{(j , k)}= \frac{1}{2}(\order{\tau}{2}_{jk}+\order{\tau}{2}_{kj})$ and $\order{\tau}{2}_{[j , k]}= \frac{1}{2}(\order{\tau}{2}_{jk}-\order{\tau}{2}_{kj})$.  From last equation, the higher-derivative torsional
terms can be written as
\begin{eqnarray}
X_1 &=& - \partial_0 T \partial_0 T + \partial_i T \partial_i T + \mathcal{O}(9),\label{X1_equ_order}\\
X_2 &=&  - \partial_0 \partial_0 T + \partial_i \partial_i T + \mathcal{O}(5) \label{X2_equ_order}.
\end{eqnarray}
Since the lowest order of velocity for $T$ is fourth order (Eq. \eqref{T_equ_order}) we can deduce that the lowest order of $X_1$ is eighth order and the lowest order of $X_2$ is  fourth order.

Also, the energy-momentum tensor expressed in term of the relevant velocity orders is given by
\begin{subequations}\label{eqn:energymomentum}
\begin{align}
\mathcal{T}_{00} &= \rho\left(1 + \Pi + v^2 - 2\order{\tau}{2}_{00}\right) + \mathcal{O}(6)\,,\\
\mathcal{T}_{0j} &= -\rho v_j + \mathcal{O}(5)\,,\\
\mathcal{T}_{ij} &= \rho v_iv_j + p\delta_{ij} + \mathcal{O}(6)\,.
\end{align}
\end{subequations}
These are all formulas which will be necessary for the post-Newtonian expansion of the field equations. We will proceed with this expansion and their solution in the following section.

\section{Field equations}\label{sec:solution}
To obtain the expression of the post-Newtonian parameters for the class of theories analyzed in this paper,  we need to  expand the field equations to each velocity order (up to fourth order)  and solve them using the post-Newtonian approximation. Following \cite{Ualikhanova2019} we also assume a generic ansatz for the perturbative terms of the tetrad field, which consists in assuming that they can be expressed as linear combinations of  constant coefficients  and post-Newtonian potentials. 

\subsection{Zeroth velocity order}\label{ssec:order0}

This order represents the background solution of the vacuum field equations. First, the energy-momentum tensor at the zeroth velocity order is null. This fact is deduced from the expansion in the Eq. \eqref{eqn:energymomentum}. Therefore, it only remains to solve the zero order of the field equations \eqref{FEquations} by introducing in them the assumed background expressions for the tetrad. Since all the terms within the expressions for $\order{E}{0}_{00}$ and $\order{E}{0}_{ij}$ are proportional to $F (0,0,0)$, this automatically leads to
\begin{equation}
\order{E}{0}_{00} = 0\,, \quad
\order{E}{0}_{ij} = 0.
\end{equation}
This is because to fulfil the post-Newtonian approximation, the function $F(T,\left(\nabla{T}\right)^2,\Box {T})$ must satisfy 
\begin{equation}
    F(0,0,0) = 0 \label{background-mink}
\end{equation}
since a flat Minkowski space is considered as the background. This condition shows a limitation of the PPN formalism given that it cannot be applied to several of the more complicated theories of gravity which are currently being used in cosmology.

\subsection{Analysis of higher orders of velocity}\label{ssec:orderx}
Before studying the higher orders  of velocity we will analyze some terms of equation \eqref{FEquations}. First, to evaluate Eq. \eqref{background-mink}, we propose the following ansatz:

\begin{eqnarray}\label{F_general}
F(T, X_1, X_2) &=& T + \sum_{n_1}\alpha_{n_1} T^{n_1} + \sum_{n_2}\alpha_{n_2} X_{2}^{n_2} \nonumber\\
&+& \sum_{n_3}\alpha_{n_3} X_{1}^{n_3}   \nonumber\\
&+&  \sum_{m_1,m_2}\alpha_{m_1,m_2} T^{m_1} X_{2}^{m_2} \nonumber\\
&+& \sum_{m_3,m_4}\alpha_{m_3,m_4} T^{m_3} X_{1}^{m_4}   \nonumber\\
&+& \sum_{m_5,m_6}\alpha_{m_5,m_6} X_{2}^{m_5} X_{1}^{m_6} \nonumber\\
&+& \sum_{m_7,m_8,m_9}\alpha_{m_7,m_8,m_9} T^{m_7}X_{2}^{m_8} X_{1}^{m_9},\nonumber\\
&&
\end{eqnarray}
where values $n_i$ and $m_i$ are greater or equal to 1 to avoid divergences. Besides, $n_2$ must be restricted to $n_2 \neq 1$, as $ \Box {T}$ is a boundary term.  

Now, using the order of velocity found in equation \eqref{X1_equ_order}, we examine the higher-derivative torsional term $X_1$ and its partial derivatives 
\begin{eqnarray}
F_{,X_{1}}(0,0,0) \frac{\partial X_{1}}{\partial e^{A}_{~\rho}} &\sim&	\order{\tau}{8}_{\mu\nu}, \\
F_{X_{1}} (0,0,0) \frac{\partial 
{X_{1}}}{\partial{\partial_{\mu} e^{A}_{~\rho}}} &\sim&	\order{\tau}{6}_{\mu\nu}, \\
F_{,X_{1}}(0,0,0) \frac{\partial X_{1}}{\partial{\partial_{\mu}\partial_{\nu} e^{A}_{~\rho}}} &\sim&	\order{\tau}{6}_{\mu\nu}, 
\end{eqnarray}
and we conclude that the  last terms are not of interest in post-Newtonian approximation. Next, we study terms related to $X_2$ (details in Appendix \ref{Appendix1}),
\begin{eqnarray}
F_{,X_{2}}(0,0,0) \frac{\partial X_{2}}{\partial e^{A}_{~\rho}} &\sim& F_{,X_{2}}(0,0,0) \order{\tau}{4}_{\mu\nu} = 0, \\
F_{,X_{2}}(0,0,0) \frac{\partial X_{2}}{\partial \partial_\mu e^{A}_{~\rho}} &\sim& F_{,X_{2}}(0,0,0) \order{\tau}{2}_{\mu\nu} = 0, \\
F_{,X_{2}}(0,0,0) \frac{\partial X_{2}}{\partial \partial_\mu\partial_\nu e^{A}_{~\rho}} &\sim& F_{,X_{2}}(0,0,0) \order{\tau}{2}_{\mu\nu} = 0, \\
F_{,X_{2}}(0,0,0) \frac{\partial X_{2}}{\partial \partial_\lambda\partial_\mu\partial_\nu e^{A}_{~\rho}} &\sim& F_{,X_{2}}(0,0,0) \order{\tau}{2}_{\mu\nu} = 0.
\end{eqnarray}
and we find that these terms are null, due to the fact that for Eq. \eqref{F_general}, $F_{,X_{2}}(0,0,0)=0$. Although we have proposed an $F (T,(\nabla T)^2, \Box{T})$ as general as possible, there are particular cases where $F (0,0,0) = 0$ but $F_{,X_{2}}(0,0,0)$ is not null (for example, models with terms such as $\Box{T} e^{T}$ \footnote{Actually, $F_{,X_{2}}(0,0,0) \neq 0$ for any function $F = \Box{T} f(T, (\nabla T)^2)$ where $f(0,0) \neq 0$.})  and the corresponding contribution must be considered. However, the analysis of these specific cases is beyond the goal of this paper.
Hence, in post-Newtonian approximation, equation \eqref{F_E_SP} reduces to

\begin{eqnarray}
&& E_{\mu \nu}\equiv F_{,T} G_{\mu \nu}+\frac{1}{4}g_{\mu \nu} (F-F_{,T}T)+\frac{\kappa^2}{2} {\mathcal{T}^{(m)}}_{\mu \nu} = 0. \nonumber\\
&&
\label{FEquations_red}
\end{eqnarray} This field equation is symmetric since the action \eqref{action0} becomes local Lorentz invariant up to fourth velocity order in the post-Newtonian expansion. Thus, we are now dealing with the usual ten degrees of freedom of curvature-based gravity theories.  Nevertheless, by analysing a higher order than fourth in Eq. \eqref{F_E_SP}, it is observed that the contributions coming from both the non-linear torsion terms and the higher-derivative torsional terms (including the effects of the additional DOF present in generic modifications of TG due to local Lorentz violation  \cite{Sotiriou:2010mv,Li:2010cg}) can contribute to the field equations from the sixth velocity order and then beyond the first PPN approximation \cite{Wu:2021ykd}. 
Furthermore, if the condition of a static vacuum background is relaxed, these contributions can  arise at lower velocity order. For instance, the idea of a dynamical background (FRW metric) has been realized in the recently proposed parameterized post-Newtonian cosmology (PPNC) framework \cite{Sanghai:2016tbi,Anton:2021vsr}.


\subsection{Second velocity order}\label{ssec:order2}
Next, we expand the field equations in terms of  the tetrad perturbations at second velocity order:
\begin{eqnarray}\label{eqn:trfieldeqn_order2}
\order{E}{2}_{00} &=&- 2 f_{T}\order{\tau}{2}_{i[i,j]j}-\kappa^2\rho, 
\nonumber\\
\order{E}{2}_{ij} &=& f_{T} \order{\tau}{2}_{j[k,i]k} + f_{T} \left( \order{\tau}{2}_{i[k,j]k} + \order{\tau}{2}_{k[j,i]k} \right) \nonumber\\
&-& f_{T} \left[ 2 \order{\tau}{2}_{k[k,i]j} - \order{\tau}{2}_{00,ij} + \left( \order{\tau}{2}_{00,kk} + 2\order{\tau}{2}_{k[l,k]l} \right) \delta_{ij} \right] \,,\nonumber\\
&&
\end{eqnarray}
where we have set out the constant $f_{T}=F_{,T}(0,0,0)$ and $\tau_{i[j,k]l} =\frac{1}{2}(\tau_{ij,kl}-\tau_{ik,jl})$.

Defining $U$ and $U_{ij}$ as the postnewtonian functionals of the matter variables, it is possible to obtain their relations with the matter variables such that:
\begin{equation}\label{eqn:PPNpotential_U}
\triangle U=-4\pi\rho\,, \quad \quad U_{ij}=\chi_{,ij} + U \delta_{ij} \quad \quad \nabla^2\chi=-2U,
\end{equation}
in which $\nabla^2 = \delta^{ij}\partial_i \partial_j$ refers to the spatial Laplace operator of the background metric and $\chi$ is the superpotential defined in \cite{Will2014}. We also assume as usual in the PPN formalism the following ansatz
\begin{align}\label{eqn:sol_tau2}
\order{\tau}{2}_{00} &= a_1 U,\nonumber\\
\order{\tau}{2}_{ij} &= a_2 U \delta_{ij} + a_3 U_{ij}\,,
\end{align}
to obtain the field equations at second order
\begin{align}\label{eqn:trfieldeqn_order2_2}
\order{E}{2}_{00} &= -\left[\kappa^2 - 8\pi(a_2+a_3)f_{T} \right]\rho, 
\\
\order{E}{2}_{ij} &= f_{T}\left( a_1 - a_2 - a_3 \right) \left( 4\pi\delta_{ij}\rho + U_{,ij} \right) \,.
\end{align}
Parameters $a_i$ are constant coefficients, which are determined from the solutions of the field equations and imposing gauge conditions. In the standard PPN gauge, the spatial part of the metric is diagonal and isotropic, consequently $a_3=0$, since  $\order{g}{2}_{ij}$ should be only proportional to $U\delta_{ij}$. Therefore, we solve the last system for $a_1$ and $a_2$ landing on
\begin{align}\label{eqn:sol_tau2_2}
a_1 &= \frac{\kappa^2}{8\pi f_{T}},\nonumber\\
a_2 &= \frac{\kappa^2}{8\pi f_{T}}\,.
\end{align}

\subsection{Third velocity order}\label{ssec:order3}
We proceed with the expansion at the third velocity order of the field equations where the only non vanishing terms are given by
\begin{eqnarray}\label{eqn:ei03a}
\order{E}{3}_{0i}  &=& f_{T}\left[ \dfrac{1}{2}\left(\order{\tau}{2}_{ij,0j}-\order{\tau}{3}_{i0,jj}+ \order{\tau}{2}_{ji,0j}-\order{\tau}{3}_{j0,ij}+2\order{\tau}{3}_{0[j,i]j}\right)\right. \nonumber\\
&-& \left.\order{\tau}{2}_{jj,0i}+\order{\tau}{3}_{j0,ij} \right] + \kappa^2 \rho v_i \,,
\\
\order{E}{3}_{i0}  &=& f_{T} \left[ \dfrac{1}{2}\left(2\order{\tau}{3}_{0[j,i]j}-\order{\tau}{2}_{00,0i}+ 2\order{\tau}{3}_{[j|0|,i]j}+2\order{\tau}{2}_{[ij],0j}-\order{\tau}{2}_{00,0i}\right)\right. \nonumber\\
&-& \left. 2\order{\tau}{2}_{j[j,|0|i]}+\order{\tau}{2}_{00,0i}\right] + \kappa^2 \rho v_i \,. \label{eqn:ei03b}
\end{eqnarray}
It should be noted that the tetrad perturbations at third velocity order $\order{\tau}{3}_{0i}$ and  $\order{\tau}{3}_{i0}$ transform as vectors under spatial rotations. Therefore, the following ansatz is assumed:
\begin{equation}\label{eqn:ansatz3}
\order{\tau}{3}_{i0} = \order{\tau}{3}_{0i} = b_1V_i+b_2W_i,
\end{equation}
being $V_i$ and $W_i$  postnewtonian functions of the matter variables defined as follows:
\begin{equation}\label{eqn:PPNpotentials_ViandWi}
\triangle V_i=-4\pi\rho v_i\,, \qquad \triangle W_i=-4\pi\rho v_i+2 U_{,0i}\,,
\end{equation}
where $b_i$ are constant parameters that are established in the same way as the $a_i$ coefficients in the subsection \ref{ssec:order3}. Rewriting expressions \eqref{eqn:ei03a} and \eqref{eqn:ei03b}
\begin{align}
\order{E}{3}_{0i}=\order{E}{3}_{i0}&= \left[\kappa^2+4\pi f_{T} (b_1+b_2) \right]\left(\rho v_i-\frac{U_{,0i}}{4\pi}\right)\,,\label{eqn:feansi03}
\end{align}
we are able to solve the system and obtain:
\begin{align}
b_1&= -b_0 -\frac{\kappa^2}{4\pi f_T} \,,\label{eqn:feansi03_2}\\
b_2&= b_0\,\label{eqn:feans00i3}.
\end{align}
leaving $b_0$ as a parameter to be determined in the next subsection instead of setting the gauge and thus having another equation \cite{Ualikhanova2019,Flathman2020}.

\subsection{Fourth velocity order}\label{ssec:order4}
Finally, we expand the field equations at fourth velocity order, and we obtain the traces for

\begin{eqnarray}\label{eqn:400order}
    \order{E}{4}_{00} &=& -\frac{f_T}{2}\left[ -\order{\tau}{2}_{ij,k}\order{\tau}{2}_{i[k,j]}+\order{\tau}{2}_{ij,k}\left(\order{\tau}{2}_{k[j,i]}+\order{\tau}{2}_{j[i,k]}\right) \right. \nonumber\\
&+& \left. \order{\tau}{2}_{ij,i}\order{\tau}{2}_{kj,k}+\order{\tau}{2}_{ii,j}\order{\tau}{2}_{kk,j}+ 2\order{\tau}{2}_{ij,i}\order{\tau}{2}_{jk,k}\right]\nonumber\\
&-&2 f_T \left[\order{\tau}{4}_{i[i,j]j}+\order{\tau}{2}_{ij,k}\order{\tau}{2}_{j[k,i]}+2\order{\tau}{2}_{00}\order{\tau}{2}_{i[j,i]j}-\order{\tau}{2}_{ii,j}\order{\tau}{2}_{(jk),k}\right. \nonumber\\
&+& \left.\order{\tau}{2}_{ij}\left(\order{\tau}{2}_{j[k,i]k}+\order{\tau}{2}_{k(i,j)k}-\order{\tau}{2}_{kk,ij}\right)\right] \nonumber\\
&-&\kappa^2 \rho v^2 - \kappa^2 \rho \Pi,
\end{eqnarray}
and
\begin{eqnarray}\label{eqn:4iiorder}
\order{E}{4}_{ii}&=&-2f_{T}\left[\order{\tau}{4}_{00,ii}-\order{\tau}{3}_{0i,0i}-\order{\tau}{2}_{00,i}\order{\tau}{2}_{ij,j}+\order{\tau}{2}_{ii}\order{\tau}{2}_{jk,jk}\right. \nonumber\\
&-& \left.\order{\tau}{2}_{ij}\order{\tau}{2}_{jk,ik}+\order{\tau}{2}_{ji}\order{\tau}{2}_{ij,kk}-\order{\tau}{2}_{kk}\order{\tau}{2}_{ii,jj}+\order{\tau}{2}_{00,ii}\left(\order{\tau}{2}_{00}+\order{\tau}{2}_{jj}\right)\right]\nonumber\\
&-&2f_{T}\left[\order{\tau}{2}_{i[i,j]}\order{\tau}{2}_{jk,k}-\order{\tau}{4}_{i[i,j]j}+\order{\tau}{2}_{ii,00}-\order{\tau}{3}_{i0,i0}\right. \nonumber\\
&+& \left. 2\order{\tau}{2}_{00,i}\order{\tau}{2}_{j[j,i]}+2\order{\tau}{2}_{ij}\left(\order{\tau}{2}_{kk,ij}-\order{\tau}{2}_{k(i,j)k}\right)\right]  \nonumber\\
&+&\frac{1}{4}f_{T}\left[2\order{\tau}{2}_{ik}\order{\tau}{2}_{ij,jk}+2\order{\tau}{2}_{kj,i}\order{\tau}{2}_{ki,j}+\order{\tau}{2}_{ij,k}\left(\order{\tau}{2}_{ij,k}-3\order{\tau}{2}_{ik,j}\right)\right. \nonumber\\
&+& \left. \order{\tau}{2}_{ij,k}\order{\tau}{2}_{kj,i}+2\order{\tau}{2}_{ij}\order{\tau}{2}_{ik,jk}\right]+ 3f_{T}\order{\tau}{2}_{(ij)}\order{\tau}{2}_{00,ij}\nonumber\\
&-&\frac{1}{2}f_{T}\left[\order{\tau}{2}_{ii,j}\left(2\order{\tau}{2}_{kj,k}-\order{\tau}{2}_{kk,j}\right)-\order{\tau}{2}_{ij,i}\order{\tau}{2}_{kj,k}-\order{\tau}{2}_{ij,k}\order{\tau}{2}_{jk,i}\right] \nonumber\\
&-&\frac{3}{4}f_{T}\order{\tau}{2}_{ij,k}\order{\tau}{2}_{ji,k} -f_{T}\order{\tau}{2}_{ij}\order{\tau}{2}_{ij,kk}\nonumber\\
&-& 3 \kappa^2 p - \kappa^2 \rho v^2 \,.
\end{eqnarray}

Next, we note that the relevant tetrad perturbation at fourth velocity order $\order{\tau}{4}_{00}$ behaves as a scalars under spatial rotations and therefore we consider the following ansatz:
\begin{equation}
\order{\tau}{4}_{00} = c_1 \Phi_1 + c_2 \Phi_2 + c_3 \Phi_3 + c_4 \Phi_4 + c_5 U^2,
\end{equation}
being $c_i$ constant coefficients (similar to $a_i$ and $b_i$) while $\Phi_i$ represent the typical PPN potentials defined by
\begin{eqnarray}
\nabla^2\Phi_1&=& -4\pi\rho v^2\,, \qquad
\nabla^2\Phi_2= -4\pi\rho U\,, \qquad \nonumber\\
\nabla^2\Phi_3&=& -4\pi\rho \Pi\,, \qquad
\nabla^2\Phi_4= -4\pi p\,.
\end{eqnarray}
Replacing the latter ansatz in Eqs. \eqref{eqn:400order} and  \eqref{eqn:4iiorder} we obtain:
\begin{eqnarray}\label{eqn:feans4}
\order{E}{4}_{00} + \order{E}{4}_{ii}
&=&-2 f_{T}\left\{2b_0U_{,00}+4\pi[c_1\rho v^2+(c_2+2c_5)\rho U\right. \nonumber\\
&+& \left.c_3\rho\Pi+c_4 p]-2c_5U_{,i}U_{,i}\right\} \\
&-&\frac{\kappa^2}{4\pi}\left(U_{,00}-\frac{\kappa^2\rho U}{2 f_{T}}\right)+3\kappa^2p+2\kappa^2\rho v^2 \nonumber\\ &+&\kappa^2\left(\rho\Pi+\frac{\kappa^2}{32\pi^2}\frac{U_{,i}U_{,i}}{f_{T}}\right)\,.
\end{eqnarray}
In order to avoid any violation of the standard PPN gauge, the coefficient that follows $U,_{00}$ must be null (in addition, it does not correspond to any term of the ansatz proposed in this subsection). On the other hand, it is also necessary that the coefficients in front of the terms $\rho U$, $p$, $\rho \Pi$, $\rho v^2$, $U,_i U,_i$ also vanish. For this, the following relations have to be fulfilled:
\begin{eqnarray}
b_0 = -\frac{\kappa^2}{16 \pi f_{T}}, \nonumber \\
c_1 = \frac{\kappa^2}{4 \pi f_{T}}, \nonumber \\
c_2=\frac{\kappa^4}{32 \pi^2 f_{T}^2}, \nonumber \\
c_3 = \frac{\kappa^2}{8 \pi f_{T}}, \nonumber \\
c_4 = \frac{3\kappa^2}{8 \pi f_{T}}, \nonumber \\
c_5 = -\frac{\kappa^4}{128 \pi^2 f_{T}^2}. 
\end{eqnarray}

\section{PPN metric and Parameters}\label{sec:metpar}
In this section, we compute the metric from the results of the previous section and obtain the PPN  parameters for the $F(T,(\nabla T)^2,\Box{T})$ generalized theories.

Thanks to the coefficients computed in section \ref{sec:solution} it is possible to calculate the tetrads. Besides, from Eq. \eqref{eqn:metricexp} the metric at different orders is obtained,
\begin{subequations}
\begin{align}
\order{g}{2}_{00} =\ &\frac{\kappa^2}{4 \pi f_{T}} U \,, \\
\order{g}{2}_{ij} =\ &\frac{\kappa^2}{4 \pi  f_{T}} U\delta_{ij} \,,\\
\order{g}{3}_{0i} =\ &-\frac{\kappa^2}{8 \pi  f_{T}}\left(\frac{7}{2}V_i+\frac{1}{2}W_i\right) \,,\\
\order{g}{4}_{00} =\ & \frac{\kappa^2}{8 \pi  f_{T}} \left( -\frac{\kappa^2}{4 \pi  f_{T}} U^2 + 4 \Phi_1 + \frac{\kappa^2}{2 \pi  f_{T}} \Phi_2 + 2 \Phi_3 + 6 \Phi_4 \right).
\end{align}
\end{subequations}
Then, assuming that the gravitational constant $G$ for this type of theories is given by,
\begin{equation}
    G=\frac{\kappa^2}{8 \pi  f_{T}}=1,
\end{equation}
the metric can be written as ,
\begin{eqnarray}
    g_{00} &=& -1 + 2 U -2 U^2 + 4 \Phi_1 + 4 \Phi_2 + 2 \Phi_3 + 6 \Phi_4 ,\\
    g_{0i} &=& -\frac{7}{2}V_i-\frac{1}{2}W_i ,\\
    g_{ij} &=& 1 + 2 U\delta_{ij} .
\end{eqnarray}
Finally, comparing the metric attained  above with the standard PPN form of the metric\cite{will1993,Will2014,Will2018}, we are able to read the PPN coefficients  
\begin{align}\label{eqn:ppnpv}
\alpha_1=\alpha_2=\alpha_3=\zeta_1 &= \zeta_2=\zeta_3=\zeta_4=\xi = 0,\\
\gamma &=\beta = 1,
\end{align}
Our results show  that for $F(T,(\nabla T)^2,\Box{T})$  theories  within the fourth order of the PPN formalism    there is no violation of the total energy-momentum conservation, nor the effects of the preferred frame or the preferred location are relevant enough. In this way these theories can be considered as fully conservative at least at these orders. In addition, our estimates for the $\beta$ and $\gamma$ parameters are equal to the ones obtained in General Relativity which in turn are consistent with the experimental and observational bounds \cite{will1993,Will2014,Will2018}.

\section{Concluding Remarks}\label{conclusion_f}
In the present paper we studied the parametrized  post-Newtonian (PPN) limit of higher-derivative-torsion modified teleparallel gravity theories. These latter theories \cite{Otalora:2016} constitute a new class of modified gravity theories which are constructed by adding higher-derivative torsional terms to the action of $F(T)$ gravity \cite{Bengochea:2008gz,Linder:2010py}. Higher order terms are motivated by the similar constructions based on curvature, whose origin is related to quantum corrections or to a fundamental gravitational
theory (e.g. string theory, Kaluza-Klein theory \cite{Nojiri:2010wj,Capozziello:2011et,Gasperini:1991ak}) or to quantum-gravity-like effective actions at scales closed to the Planck scale \cite{Vilkovisky:1992pb}.   In this context, torsion is associated to the Weitzenb\"{o}ck connection of teleparallel gravity  \cite{Einstein,TranslationEinstein,Early-papers1,Early-papers2,Early-papers3,Early-papers4,Early-papers5,Early-papers6,JGPereira2,AndradeGuillenPereira-00,Arcos:2005ec,Pereira:2019woq,Aldrovandi-Pereira-book,Krssak:2015rqa}. Furthermore, the PPN formalism provides a remarkable tool in studying the viability of gravity theories to fulfill the constraints imposed by local-scale observations through a set of ten parameters that have been measured with a high precision \cite{will1993,Will2014,Will2018}. 

We have started from the covariant formulation of modified teleparallel gravity by restoring the non-vanishing spin connection of the theory \cite{Krssak:2015oua, Krssak:2015rqa}. Thus, in order to obtain the PPN limit we expanded the tetrad field around the Minkowski background and found the corresponding perturbed field equations. At this point, by establishing the ansatz for the perturbed tetrad field consistent with the standard PPN spacetime metric we have clarified the count of the total number of degrees of freedom (DOF) including the six additional modes appearing in modified teleparallel gravity (MTG) due to local Lorentz symmetry breaking \cite{Sotiriou:2010mv,Li:2010cg}. In this way, by using this PPN expansion of the tetrad field we calculated the relevant geometrical quantities at hand, as for instance the torsion scalar, and the higher-derivative torsional terms up to fourth velocity order. With these results we have shown that the torsion scalar is fourth velocity order, as well as the higher-derivative torsional terms (e.g. $\Box{T}$, $(\nabla{T})^2$) which are fourth and eighth order respectively. Therefore, we have shown that the contributions to the perturbed field equations originated from the modifications to teleparallel gravity  (products of non-linear torsion terms or due to higher-derivative torsional terms, including the effects of additional perturbative modes) can appear explicitly only from the sixth velocity order, that is to say, beyond the PPN formalism. Therefore, a second post-Newtonian (2PN) order approximation should be considered to study these contributions \cite{Wu:2021ykd}. Consequently, in the traditional PPN formalism it is not possible to find for the theories studied here, any deviations from the PPN parameters with respect to the GR predictions (consistent with the experimental and observational bounds).

Finally, it is important to note that the PPN formalism relies on the asymptotical 
flatness and slow-motion assumptions which are not valid on larger scales as the cosmological one. At the same time, with the motivation to extend the success of the PPN formalism to cosmological scales, and to encompass a larger class of theories of gravity and dark energy models as possible, an attempt to construct a parameterized post-Newtonian cosmology (PPNC) has been performed in Refs. \cite{Sanghai:2016tbi,Anton:2021vsr}. This new formalism takes into account the time dependence of the cosmological quantities linked to the large scale expansion, as well as it is still valid in the presence of non-linear structures and consistent with the PPN limit. In this sense,  we have  shown that the consequences of non-linear torsion terms, higher-derivative torsion terms, including the effects of the additional perturbative modes present in MTG, can also become significant if the condition of the static vacuum background is relaxed. Thus, in order to study the new observational imprints predicted by these theories a new analysis within the PPNC framework is mandatory.
\section{Acknowledgments}
M. Gonzalez-Espinoza acknowledges support from Proyecto Postdoctorado $2021$ VRIEA-PUCV. G. Otalora acknowldeges DI-VRIEA for financial support through Proyecto Postdoctorado $2020$ VRIEA-PUCV. L. Kraiselburd and S. Landau are supported by the National Agency forthe  Promotion  of  Science  and  Technology  (ANPCYT)of  Argentina  grant  PICT-2016-0081; CONICET grant PIP 11220200100729CO and  grants G157 and G175 from UNLP.

\bibliographystyle{spphys}       
\bibliography{bio}   

\begin{widetext}
\section{Appendix: partial derivatives of the higher-derivative torsional terms}\label{Appendix1}
In this appendix, we show the partial derivatives of the higher-derivative torsional term $X_2$ using CADABRA \cite{Peeters:2007wn,Peeters:2018dyg}. First, the partial derivative of $X_2$ with respect to $e^{A}_{~\gamma}$ can be expressed as
\begin{eqnarray}
\dfrac{\partial X_2}{\partial e^A_{~\gamma}} &=& \eta^{\gamma \mu_{1}} \delta_{A}^{~\nu_{1}}\Bigg[ \frac{1}{2}\partial_{\alpha}\left( \order{\tau}{2}_{\mu1 \zeta}\right) \partial_{\nu1}\,^{\beta}\,_{\beta}\left( \order{\tau}{2}^{\alpha \zeta}\right)  +\frac{1}{4}\partial_{\alpha}\left( \order{\tau}{2}_{\nu1}\,^{\mu}\right) \partial^{\beta}\,_{\beta \mu}\left( \order{\tau}{2}_{\mu1}\,^{\alpha}\right)  +\frac{1}{2}\partial_{\alpha}\left( \order{\tau}{2}^{\beta}\,_{\zeta}\right) \partial_{\mu1 \nu1 \beta}\left( \order{\tau}{2}^{\alpha \zeta}\right) 
\nonumber\\ && +\frac{5}{4}\partial_{\mu1}\left( \order{\tau}{2}^{\beta}\,_{\zeta}\right) \partial^{\alpha}\,_{\alpha \beta}\left( \order{\tau}{2}_{\nu1}\,^{\zeta}\right)  +\frac{5}{4}\partial_{\nu1}\left( \order{\tau}{2}^{\beta}\,_{\zeta}\right) \partial^{\alpha}\,_{\alpha \beta}\left( \order{\tau}{2}_{\mu1}\,^{\zeta}\right)   - \frac{3}{2}\partial^{\alpha}\left( \order{\tau}{2}_{\mu1 \zeta}\right) \partial_{\alpha}\,^{\beta}\,_{\beta}\left( \order{\tau}{2}_{\nu1}\,^{\zeta}\right) \nonumber\\ &&  - \frac{3}{4}\partial^{\alpha}\left( \order{\tau}{2}_{\nu1}\,^{\mu}\right) \partial_{\alpha}\,^{\beta}\,_{\beta}\left( \order{\tau}{2}_{\mu1 \mu}\right)   - \frac{5}{2}\partial^{\alpha}\left( \order{\tau}{2}^{\beta}\,_{\zeta}\right) \partial_{\mu1 \nu1 \alpha}\left( \order{\tau}{2}_{\beta}\,^{\zeta}\right)   - \frac{5}{4}\partial_{\mu1}\left( \order{\tau}{2}^{\beta}\,_{\zeta}\right) \partial_{\nu1}\,^{\alpha}\,_{\alpha}\left( \order{\tau}{2}_{\beta}\,^{\zeta}\right) \nonumber\\ &&  - \frac{5}{4}\partial_{\nu1}\left( \order{\tau}{2}^{\beta}\,_{\zeta}\right) \partial_{\mu1}\,^{\alpha}\,_{\alpha}\left( \order{\tau}{2}_{\beta}\,^{\zeta}\right)   - \frac{1}{2}\partial_{\zeta}\left( \order{\tau}{2}_{\mu1 \beta}\right) \partial_{\nu1}\,^{\alpha}\,_{\alpha}\left( \order{\tau}{2}^{\beta \zeta}\right)   - \frac{1}{4}\partial_{\nu1}\left( \order{\tau}{2}^{\beta}\,_{\mu}\right) \partial^{\alpha}\,_{\alpha \beta}\left( \order{\tau}{2}_{\mu1}\,^{\mu}\right) 
\nonumber\\ && - \frac{1}{2}\partial_{\zeta}\left( \order{\tau}{2}^{\alpha}\,_{\beta}\right) \partial_{\mu1 \nu1 \alpha}\left( \order{\tau}{2}^{\beta \zeta}\right)   - \frac{1}{4}\partial_{\zeta}\left( \order{\tau}{2}_{\mu1}\,^{\beta}\right) \partial^{\alpha}\,_{\alpha \beta}\left( \order{\tau}{2}_{\nu1}\,^{\zeta}\right)   - \frac{3}{4}\partial_{\zeta}\left( \order{\tau}{2}_{\nu1}\,^{\beta}\right) \partial^{\alpha}\,_{\alpha \beta}\left( \order{\tau}{2}_{\mu1}\,^{\zeta}\right)  
\nonumber\\ && +\frac{1}{2}\partial_{\zeta}\left( \order{\tau}{2}_{\mu1 \beta}\right) \partial^{\alpha}\,_{\alpha}\,^{\beta}\left( \order{\tau}{2}_{\nu1}\,^{\zeta}\right)  +\frac{1}{4}\partial_{\nu1}\left( \order{\tau}{2}^{\mu}\,_{\beta}\right) \partial^{\alpha}\,_{\alpha}\,^{\beta}\left( \order{\tau}{2}_{\mu1 \mu}\right)  +\frac{1}{2}\partial_{\zeta}\left( \order{\tau}{2}^{\beta}\,_{\alpha}\right) \partial_{\mu1 \nu1}\,^{\alpha}\left( \order{\tau}{2}_{\beta}\,^{\zeta}\right) 
\nonumber\\ && +\frac{1}{4}\partial_{\zeta}\left( \order{\tau}{2}_{\mu1}\,^{\beta}\right) \partial_{\nu1}\,^{\alpha}\,_{\alpha}\left( \order{\tau}{2}_{\beta}\,^{\zeta}\right)  %
+\frac{3}{4}\partial_{\zeta}\left( \order{\tau}{2}_{\nu1}\,^{\beta}\right) \partial_{\mu1}\,^{\alpha}\,_{\alpha}\left( \order{\tau}{2}_{\beta}\,^{\zeta}\right)   - \frac{3}{4}\partial^{\alpha}\left( \order{\tau}{2}_{\nu1}\,^{\zeta}\right) \partial_{\alpha}\,^{\beta}\,_{\beta}\left( \order{\tau}{2}_{\mu1 \zeta}\right) 
\nonumber\\ && - \frac{1}{2}\partial^{\alpha}\left( \order{\tau}{2}^{\zeta}\,_{\kappa}\right) \partial_{\mu1 \nu1 \alpha}\left( \order{\tau}{2}_{\zeta}\,^{\kappa}\right)   - \frac{1}{4}\partial_{\mu1}\left( \order{\tau}{2}^{\zeta}\,_{\kappa}\right) \partial_{\nu1}\,^{\alpha}\,_{\alpha}\left( \order{\tau}{2}_{\zeta}\,^{\kappa}\right)   - \frac{1}{4}\partial_{\nu1}\left( \order{\tau}{2}^{\zeta}\,_{\kappa}\right) \partial_{\mu1}\,^{\alpha}\,_{\alpha}\left( \order{\tau}{2}_{\zeta}\,^{\kappa}\right) 
\nonumber\\ && +\frac{1}{4}\partial_{\alpha}\left( \order{\tau}{2}_{\nu1}\,^{\zeta}\right) \partial^{\beta}\,_{\beta \zeta}\left( \order{\tau}{2}_{\mu1}\,^{\alpha}\right)  +\frac{1}{2}\partial_{\alpha}\left( \order{\tau}{2}^{\zeta}\,_{\kappa}\right) \partial_{\mu1 \nu1 \zeta}\left( \order{\tau}{2}^{\alpha \kappa}\right)  +\frac{1}{4}\partial_{\mu1}\left( \order{\tau}{2}^{\zeta}\,_{\kappa}\right) \partial^{\alpha}\,_{\alpha \zeta}\left( \order{\tau}{2}_{\nu1}\,^{\kappa}\right) \nonumber\\ && +\frac{1}{4}\partial_{\nu1}\left( \order{\tau}{2}^{\zeta}\,_{\kappa}\right) \partial^{\alpha}\,_{\alpha \zeta}\left( \order{\tau}{2}_{\mu1}\,^{\kappa}\right)  +\frac{3}{4}\partial_{\nu1}\left( \order{\tau}{2}_{\beta}\,^{\zeta}\right) \partial^{\alpha}\,_{\alpha}\,^{\beta}\left( \order{\tau}{2}_{\mu1 \zeta}\right)  +\frac{1}{2}\partial_{\zeta}\left( \order{\tau}{2}_{\alpha}\,^{\kappa}\right) \partial_{\mu1 \nu1}\,^{\alpha}\left( \order{\tau}{2}^{\zeta}\,_{\kappa}\right)  \nonumber\\ && +\frac{1}{4}\partial_{\zeta}\left( \order{\tau}{2}_{\mu1}\,^{\kappa}\right) \partial_{\nu1}\,^{\alpha}\,_{\alpha}\left( \order{\tau}{2}^{\zeta}\,_{\kappa}\right)  +\frac{1}{4}\partial_{\zeta}\left( \order{\tau}{2}_{\nu1}\,^{\kappa}\right) \partial_{\mu1}\,^{\alpha}\,_{\alpha}\left( \order{\tau}{2}^{\zeta}\,_{\kappa}\right)   - \frac{1}{4}\partial_{\nu1}\left( \order{\tau}{2}_{\beta}\,^{\zeta}\right) \partial^{\alpha}\,_{\alpha \zeta}\left( \order{\tau}{2}_{\mu1}\,^{\beta}\right) 
\nonumber\\ && - \frac{1}{2}\partial_{\zeta}\left( \order{\tau}{2}_{\alpha}\,^{\kappa}\right) \partial_{\mu1 \nu1 \kappa}\left( \order{\tau}{2}^{\alpha \zeta}\right)   - \frac{1}{4}\partial_{\zeta}\left( \order{\tau}{2}_{\mu1}\,^{\kappa}\right) \partial^{\alpha}\,_{\alpha \kappa}\left( \order{\tau}{2}_{\nu1}\,^{\zeta}\right)   - \frac{1}{4}\partial_{\zeta}\left( \order{\tau}{2}_{\nu1}\,^{\kappa}\right) \partial^{\alpha}\,_{\alpha \kappa}\left( \order{\tau}{2}_{\mu1}\,^{\zeta}\right) 
\nonumber\\ && -\partial^{\alpha}\left( \order{\tau}{2}_{\beta}\,^{\zeta}\right) \partial_{\mu1 \nu1 \alpha}\left( \order{\tau}{2}^{\beta}\,_{\zeta}\right)   - \frac{1}{2}\partial^{\alpha}\left( \order{\tau}{2}_{\mu1}\,^{\zeta}\right) \partial_{\alpha}\,^{\beta}\,_{\beta}\left( \order{\tau}{2}_{\nu1 \zeta}\right)   - \frac{1}{2}\partial_{\mu1}\left( \order{\tau}{2}_{\beta}\,^{\zeta}\right) \partial_{\nu1}\,^{\alpha}\,_{\alpha}\left( \order{\tau}{2}^{\beta}\,_{\zeta}\right)  %
\nonumber\\ &&
 - \frac{1}{2}\partial_{\nu1}\left( \order{\tau}{2}_{\beta}\,^{\zeta}\right) \partial_{\mu1}\,^{\alpha}\,_{\alpha}\left( \order{\tau}{2}^{\beta}\,_{\zeta}\right)  +\partial_{\alpha}\left( \order{\tau}{2}_{\beta}\,^{\zeta}\right) \partial_{\mu1 \nu1}\,^{\beta}\left( \order{\tau}{2}^{\alpha}\,_{\zeta}\right)  +\frac{1}{2}\partial_{\alpha}\left( \order{\tau}{2}_{\mu1}\,^{\zeta}\right) \partial_{\nu1}\,^{\beta}\,_{\beta}\left( \order{\tau}{2}^{\alpha}\,_{\zeta}\right)  \nonumber\\ && +\frac{1}{2}\partial_{\alpha}\left( \order{\tau}{2}_{\nu1}\,^{\zeta}\right) \partial_{\mu1}\,^{\beta}\,_{\beta}\left( \order{\tau}{2}^{\alpha}\,_{\zeta}\right)  +\frac{1}{2}\partial_{\mu1}\left( \order{\tau}{2}_{\beta}\,^{\zeta}\right) \partial^{\alpha}\,_{\alpha}\,^{\beta}\left( \order{\tau}{2}_{\nu1 \zeta}\right)  -\partial_{\alpha}\left( \order{\tau}{2}_{\mu1 \nu1}\right) \partial^{\beta}\,_{\beta \zeta}\left( \order{\tau}{2}^{\alpha \zeta}\right) \nonumber\\ && -\partial_{\alpha}\left( \order{\tau}{2}^{\mu}\,_{\mu}\right) \partial_{\nu1}\,^{\beta}\,_{\beta}\left( \order{\tau}{2}_{\mu1}\,^{\alpha}\right)  -2\partial_{\alpha}\left( \order{\tau}{2}^{\beta}\,_{\beta}\right) \partial_{\mu1 \nu1 \zeta}\left( \order{\tau}{2}^{\alpha \zeta}\right)  -2\partial_{\mu1}\left( \order{\tau}{2}^{\beta}\,_{\beta}\right) \partial^{\alpha}\,_{\alpha \zeta}\left( \order{\tau}{2}_{\nu1}\,^{\zeta}\right) \nonumber\\ && -2\partial_{\nu1}\left( \order{\tau}{2}^{\beta}\,_{\beta}\right) \partial^{\alpha}\,_{\alpha \zeta}\left( \order{\tau}{2}_{\mu1}\,^{\zeta}\right)  +2\partial^{\alpha}\left( \order{\tau}{2}_{\mu1 \nu1}\right) \partial_{\alpha}\,^{\beta}\,_{\beta}\left( \order{\tau}{2}^{\mu}\,_{\mu}\right)  +2\partial^{\alpha}\left( \order{\tau}{2}^{\mu}\,_{\mu}\right) \partial_{\alpha}\,^{\beta}\,_{\beta}\left( \order{\tau}{2}_{\mu1 \nu1}\right) \nonumber\\ && +4\partial^{\alpha}\left( \order{\tau}{2}^{\beta}\,_{\beta}\right) \partial_{\mu1 \nu1 \alpha}\left( \order{\tau}{2}^{\mu}\,_{\mu}\right)  +2\partial_{\mu1}\left( \order{\tau}{2}^{\beta}\,_{\beta}\right) \partial_{\nu1}\,^{\alpha}\,_{\alpha}\left( \order{\tau}{2}^{\mu}\,_{\mu}\right)  +2\partial_{\nu1}\left( \order{\tau}{2}^{\beta}\,_{\beta}\right) \partial_{\mu1}\,^{\alpha}\,_{\alpha}\left( \order{\tau}{2}^{\mu}\,_{\mu}\right) \nonumber\\ && +\partial_{\nu1}\left( \order{\tau}{2}_{\mu1 \beta}\right) \partial^{\alpha}\,_{\alpha \zeta}\left( \order{\tau}{2}^{\beta \zeta}\right)  +\partial_{\zeta}\left( \order{\tau}{2}_{\beta}\,^{\zeta}\right) \partial_{\nu1}\,^{\alpha}\,_{\alpha}\left( \order{\tau}{2}_{\mu1}\,^{\beta}\right)  +2\partial_{\zeta}\left( \order{\tau}{2}_{\alpha}\,^{\zeta}\right) \partial_{\mu1 \nu1 \kappa}\left( \order{\tau}{2}^{\alpha \kappa}\right) \nonumber\\ && +2\partial_{\zeta}\left( \order{\tau}{2}_{\mu1}\,^{\zeta}\right) \partial^{\alpha}\,_{\alpha \kappa}\left( \order{\tau}{2}_{\nu1}\,^{\kappa}\right)  +2\partial_{\zeta}\left( \order{\tau}{2}_{\nu1}\,^{\zeta}\right) \partial^{\alpha}\,_{\alpha \kappa}\left( \order{\tau}{2}_{\mu1}\,^{\kappa}\right)  %
-\partial_{\nu1}\left( \order{\tau}{2}_{\mu1 \beta}\right) \partial^{\alpha}\,_{\alpha}\,^{\beta}\left( \order{\tau}{2}^{\mu}\,_{\mu}\right) \nonumber\\ && -\partial_{\zeta}\left( \order{\tau}{2}_{\beta}\,^{\zeta}\right) \partial^{\alpha}\,_{\alpha}\,^{\beta}\left( \order{\tau}{2}_{\mu1 \nu1}\right)  -2\partial_{\zeta}\left( \order{\tau}{2}_{\alpha}\,^{\zeta}\right) \partial_{\mu1 \nu1}\,^{\alpha}\left( \order{\tau}{2}^{\beta}\,_{\beta}\right)  -2\partial_{\zeta}\left( \order{\tau}{2}_{\mu1}\,^{\zeta}\right) \partial_{\nu1}\,^{\alpha}\,_{\alpha}\left( \order{\tau}{2}^{\beta}\,_{\beta}\right) \nonumber\\ && -2\partial_{\zeta}\left( \order{\tau}{2}_{\nu1}\,^{\zeta}\right) \partial_{\mu1}\,^{\alpha}\,_{\alpha}\left( \order{\tau}{2}^{\beta}\,_{\beta}\right)  +\frac{1}{2}\partial^{\alpha}\,_{\beta}\left( \order{\tau}{2}_{\mu1 \zeta}\right) \partial_{\nu1 \alpha}\left( \order{\tau}{2}^{\beta \zeta}\right)  +\frac{1}{4}\partial^{\alpha}\,_{\beta}\left( \order{\tau}{2}_{\nu1}\,^{\mu}\right) \partial_{\alpha \mu}\left( \order{\tau}{2}_{\mu1}\,^{\beta}\right) \nonumber\\ &&  +\frac{1}{4}\partial_{\nu1 \alpha}\left( \order{\tau}{2}^{\beta}\,_{\zeta}\right) \partial_{\mu1 \beta}\left( \order{\tau}{2}^{\alpha \zeta}\right)  +\frac{1}{4}\partial_{\mu1 \alpha}\left( \order{\tau}{2}^{\beta}\,_{\zeta}\right) \partial_{\nu1 \beta}\left( \order{\tau}{2}^{\alpha \zeta}\right)  +\frac{5}{4}\partial_{\mu1}\,^{\alpha}\left( \order{\tau}{2}^{\beta}\,_{\zeta}\right) \partial_{\alpha \beta}\left( \order{\tau}{2}_{\nu1}\,^{\zeta}\right) \nonumber\\ && +\frac{5}{4}\partial_{\nu1}\,^{\alpha}\left( \order{\tau}{2}^{\beta}\,_{\zeta}\right) \partial_{\alpha \beta}\left( \order{\tau}{2}_{\mu1}\,^{\zeta}\right)   - \frac{7}{4}\partial^{\alpha \beta}\left( \order{\tau}{2}_{\nu1}\,^{\zeta}\right) \partial_{\alpha \beta}\left( \order{\tau}{2}_{\mu1 \zeta}\right)   - \frac{3}{4}\partial^{\alpha \beta}\left( \order{\tau}{2}_{\nu1}\,^{\mu}\right) \partial_{\alpha \beta}\left( \order{\tau}{2}_{\mu1 \mu}\right) \nonumber\\ &&  - \frac{5}{2}\partial_{\nu1}\,^{\alpha}\left( \order{\tau}{2}^{\beta}\,_{\zeta}\right) \partial_{\mu1 \alpha}\left( \order{\tau}{2}_{\beta}\,^{\zeta}\right)   - \frac{5}{2}\partial_{\mu1}\,^{\alpha}\left( \order{\tau}{2}^{\beta}\,_{\zeta}\right) \partial_{\nu1 \alpha}\left( \order{\tau}{2}_{\beta}\,^{\zeta}\right)   - \frac{1}{2}\partial_{\nu1}\,^{\alpha}\left( \order{\tau}{2}^{\beta \zeta}\right) \partial_{\alpha \zeta}\left( \order{\tau}{2}_{\mu1 \beta}\right)  \nonumber\\ && - \frac{1}{4}\partial_{\nu1}\,^{\alpha}\left( \order{\tau}{2}^{\beta}\,_{\mu}\right) \partial_{\alpha \beta}\left( \order{\tau}{2}_{\mu1}\,^{\mu}\right)   - \frac{1}{4}\partial_{\mu1 \alpha}\left( \order{\tau}{2}^{\beta \zeta}\right) \partial_{\nu1 \zeta}\left( \order{\tau}{2}^{\alpha}\,_{\beta}\right)   - \frac{1}{4}\partial_{\nu1 \alpha}\left( \order{\tau}{2}^{\beta \zeta}\right) \partial_{\mu1 \zeta}\left( \order{\tau}{2}^{\alpha}\,_{\beta}\right) \nonumber\\ &&  - \frac{3}{4}\partial^{\alpha}\,_{\beta}\left( \order{\tau}{2}_{\mu1}\,^{\zeta}\right) \partial_{\alpha \zeta}\left( \order{\tau}{2}_{\nu1}\,^{\beta}\right)  %
+\frac{1}{2}\partial^{\alpha}\,_{\zeta}\left( \order{\tau}{2}_{\mu1 \beta}\right) \partial_{\alpha}\,^{\beta}\left( \order{\tau}{2}_{\nu1}\,^{\zeta}\right)  +\frac{1}{4}\partial_{\nu1}\,^{\alpha}\left( \order{\tau}{2}^{\mu}\,_{\beta}\right) \partial_{\alpha}\,^{\beta}\left( \order{\tau}{2}_{\mu1 \mu}\right) \Bigg] + \mathcal{O}(5),
\end{eqnarray}
and, the partial derivative of $X_2$ with respect to $\partial_{\rho}e^{A1}_{~~\nu1}$ can be written
\begin{eqnarray}
\dfrac{\partial X_2}{\partial \partial_{\rho1} e^{A_{1}}_{~\nu1}}&=& -\partial_{A1}\,^{\alpha}\,_{\alpha}\left( \order{\tau}{2}^{\nu1 \rho1}\right)  +\partial^{\rho1 \alpha}\,_{\alpha}\left( \order{\tau}{2}^{\nu1}\,_{A1}\right) +\partial_{A1}\,^{\alpha}\,_{\alpha}\left( \order{\tau}{2}^{\rho1 \nu1}\right)  -\partial^{\nu1 \alpha}\,_{\alpha}\left( \order{\tau}{2}^{\rho1}\,_{A1}\right)  +\partial^{\rho1 \alpha}\,_{\alpha}\left( \order{\tau}{2}_{A1}\,^{\nu1}\right) \nonumber \\\nonumber
&&-\partial^{\nu1 \alpha}\,_{\alpha}\left( \order{\tau}{2}_{A1}\,^{\rho1}\right)  +2\delta_{A1}\,^{\nu1} \partial_{A}\,^{\alpha}\,_{\alpha}\left( \order{\tau}{2}^{A \rho1}\right)  -2\delta_{A1}\,^{\nu1} \partial^{\rho1 \alpha}\,_{\alpha}\left( \order{\tau}{2}^{\beta}\,_{\beta}\right)  -2\delta_{A1}\,^{\rho1} \partial_{A}\,^{\alpha}\,_{\alpha}\left( \order{\tau}{2}^{A \nu1}\right)  
\\
&& + 2\delta_{A1}\,^{\rho1} \partial^{\nu1 \alpha}\,_{\alpha}\left( \order{\tau}{2}^{\beta}\,_{\beta}\right) + \mathcal{O}(3),  
\end{eqnarray}
Furthermore, the partial derivative of $X_2$ with respect to $\partial_{\nu1} \partial_{\tau1}e^{A_{1}}_{~\sigma1}$ reads
\begin{eqnarray}
\dfrac{\partial X_2}{\partial \partial_{\nu1 \tau1}e^{A_{1}}_{~\sigma1}} &=& -2\partial_{A1}\,^{\nu1}\left( \order{\tau}{2}^{\sigma1 \tau1}\right)  -2\partial^{\nu1 \sigma1}\left( \order{\tau}{2}^{\tau1}\,_{A1}\right)  +2\partial^{\nu1 \tau1}\left( \order{\tau}{2}^{\sigma1}\,_{A1}\right)  +2\partial_{A1}\,^{\nu1}\left( \order{\tau}{2}^{\tau1 \sigma1}\right)   \nonumber\\
&&
+2\partial^{\nu1 \tau1}\left( \order{\tau}{2}_{A1}\,^{\sigma1}\right) -2\partial^{\nu1 \sigma1}\left( \order{\tau}{2}_{A1}\,^{\tau1}\right)  +4\delta_{A1}\,^{\sigma1} \partial^{\nu1}\,_{A}\left( \order{\tau}{2}^{A \tau1}\right)  +4\delta_{A1}\,^{\tau1} \partial^{\nu1 \sigma1}\left( \order{\tau}{2}^{\alpha}\,_{\alpha}\right)   \nonumber\\
&&
-4\delta_{A1}\,^{\sigma1} \partial^{\nu1 \tau1}\left( \order{\tau}{2}^{\alpha}\,_{\alpha}\right)  -4\delta_{A1}\,^{\tau1} \partial^{\nu1}\,_{A}\left( \order{\tau}{2}^{A \sigma1}\right)+ \mathcal{O}(3),  
\end{eqnarray}

and, finally, we have obtained the partial derivative of $X_2$ with respect to $\partial_{\gamma1} \partial_{\nu1} \partial_{\tau1}e^{A_{1}}_{~\sigma1}$

\begin{eqnarray}
\dfrac{\partial X_2}{\partial \partial_{\gamma1 \nu1 \tau1}e^{A1}_{~~\sigma1}} &=& -\eta^{\gamma1 \nu1} \partial^{\sigma1}\left( \order{\tau}{2}^{\tau1}\,_{A1}\right)  +\eta^{\gamma1 \nu1} \partial^{\tau1}\left( \order{\tau}{2}^{\sigma1}\,_{A1}\right)  +\eta^{\gamma1 \nu1} \partial_{A1}\left( \order{\tau}{2}^{\tau1 \sigma1}\right)  -\eta^{\gamma1 \nu1} \partial_{A1}\left( \order{\tau}{2}^{\sigma1 \tau1}\right)
\nonumber\\ &&+\eta^{\gamma1 \nu1} \partial^{\tau1}\left( \order{\tau}{2}_{A1}\,^{\sigma1}\right)  -\eta^{\gamma1 \nu1} \partial^{\sigma1}\left( \order{\tau}{2}_{A1}\,^{\tau1}\right)  +2\delta_{A1}\,^{\tau1} \eta^{\gamma1 \nu1} \partial^{\sigma1}\left( \order{\tau}{2}^{\alpha}\,_{\alpha}\right)
\nonumber\\ && -2\delta_{A1}\,^{\sigma1} \eta^{\gamma1 \nu1} \partial^{\tau1}\left( \order{\tau}{2}^{\alpha}\,_{\alpha}\right)  -2\delta_{A1}\,^{\tau1} \eta^{\gamma1 \nu1} \partial_{A}\left( \order{\tau}{2}^{A \sigma1}\right)  +2\delta_{A1}\,^{\sigma1} \eta^{\gamma1 \nu1} \partial_{A}\left( \order{\tau}{2}^{A \tau1}\right) + \mathcal{O}(3). 
\end{eqnarray}
\end{widetext}

\end{document}